\documentclass[pra,twocolumn,aps,superscriptaddress,showpacs,longbibliography]{revtex4-1}
\usepackage{hyperref}
\hypersetup{
	colorlinks=true,
	citecolor=blue,
	linkcolor=blue,
	urlcolor=blue}
\usepackage{graphicx}
\usepackage{amsmath}
\usepackage{physics}
\usepackage{amsfonts}
\usepackage{amssymb}
\usepackage{bm}
\usepackage{siunitx}
\usepackage{xfrac}
\usepackage{braket}
\usepackage{cases}

\begin{document}
\title[]{Finite temperature spin dynamics of a two-dimensional Bose-Bose atomic mixture}

\author{Arko Roy}
\thanks{These two authors contributed equally to this work.}
\address{INO-CNR BEC Center and Universit\`a di Trento, via Sommarive 14, I-38123 Trento, Italy} 

\author{Miki Ota}
\thanks{These two authors contributed equally to this work.}
\address{INO-CNR BEC Center and Universit\`a di Trento, via Sommarive 14, I-38123 Trento, Italy} 

\author{Alessio Recati}
\address{INO-CNR BEC Center and Universit\`a di Trento, via Sommarive 14, I-38123 Trento, Italy} 
\address{
Trento Institute for Fundamental Physics and Applications, INFN, 38123 Povo, Italy}

\author{Franco Dalfovo}
\address{INO-CNR BEC Center and Universit\`a di Trento, via Sommarive 14, I-38123 Trento, Italy} 


\begin{abstract}
We examine the role of thermal fluctuations in uniform two-dimensional binary Bose mixtures of dilute ultracold atomic gases. We use a mean-field Hartree-Fock theory to derive analytical predictions for the miscible-immiscible transition. A nontrivial result of this theory is that a fully miscible phase at $T=0$ may become unstable at $T\neq0$, as a consequence of a divergent behaviour in the spin susceptibility. We test this prediction by performing numerical simulations with the Stochastic (Projected) Gross-Pitaevskii equation, which includes beyond mean-field effects. We calculate the equilibrium configurations at different temperatures and interaction strengths and we simulate spin oscillations produced by a weak external perturbation. Despite some qualitative agreement, the comparison between the two theories shows that the mean-field approximation is not able to properly describe the behavior of the two-dimensional mixture near the  miscible-immiscible transition, as thermal fluctuations smoothen all sharp features both in the phase diagram and in spin dynamics, except for temperature well below the critical temperature for superfluidity.

\end{abstract}

\maketitle


\section{Introduction}
\label{Introduction}

The study of phase-separation in two-component classical fluids is of paramount importance and the role of temperature can be rather nontrivial. Two-component (or multi-component) quantum fluids are also available ~\cite{pethick_08,pitaevskii_16} and they have been the subject of intense theoretical investigation over the last two decades~\cite{ho_96,timmermans_98,ao_98,trippenbach_2000,schaeybroeck_08, wen_12, roy_15a,ota_19}. This is supplemented by the ongoing experimental efforts with binary quantum gases~\cite{modugno_02,thalhammer_08,lercher_11,mccarron_11,pasquiou_13,papp_08,tojo_10,nicklas_11}.
In particular, the observation of two-species Bose-Einstein condensate with atoms of the same element in different hyperfine states~\cite{myatt_97,stamper_kurn_98,stenger_98,sadler_06} has received much attention because of its simplicity, yet, reveal essential kinetics
related to the transition~\cite{kawaguchi_12,stamperkurn_13}. 

The interaction in dilute gases of ultracold atoms is entirely determined by the $s$-wave scattering length, $a$, which in turn fixes the interaction coupling constant $g$ entering all relevant equations at the mean-field level. In a two-component mixture, one has two intra-component coupling constants, $g_{11}$ and  $g_{22}$, and one inter-component coupling constant $g_{12}$. The theoretical constraint for phase-separation at zero temperature is that these constants must satisfy the inequality $g_{12}^2 > g_{11}g_{22}$~\cite{pethick_08,pitaevskii_16}. However, at finite temperature, deviations from this constraint are expected to appear. Existing theoretical studies have mainly addressed quasi-one-dimensional (1D) or 3D systems employing the mean-field treatment including the Hartree-Fock~\cite{ohberg_98,shi_2000,schaeybroeck_13}, Hartree-Fock-Bogoliubov-Popov~\cite{ohberg_99,roy_15a,armaitis_15},
and Zaremba-Nikuni-Griffin formalism~\cite{lee_16,griffin_09}. Stochastic growth dynamics for quasi-1D and quasi-2D multi-component mixtures has been investigated in Refs.~\cite{su_11,liu_16}. However, there have been few finite temperature studies for 2D homogeneous Bose mixtures using beyond mean-field theory~\cite{karle_19,kobayashi_19}.  Recently, Ota \emph{et al.}~\cite{ota_19} predicted a temperature induced magnetic phase transition in an uniform 3D Bose-Bose mixture using the Popov theory. It is then natural to ask whether such a phase-transition also exists in 2D.

It is worth stressing that, in 2D Bose gases, thermal fluctuations are much more important than in 3D, as they inhibit true Bose-Einstein condensation at finite temperature, according to the Mermin-Wagner-Hohenberg theorem ~\cite{hohenberg_65,mermin_66}. Nevertheless, superfluidity still exists below the critical temperature $T_{\rm BKT}$ for the Berezinskii-Kosterlitz-Thouless (BKT) phase transition~\cite{berezinskii_72,kosterlitz_72,kosterlitz_73,kosterlitz_16}; the transition from normal gas to superfluid follows from the binding and unbinding of vortex-antivortex pairs at $T_{\rm BKT}$~\cite{prokofev_01}. Observation of such transition in the domain of ultracold quantum gases has been possible with quasi-uniform box traps~\cite{gaunt_13,chomaz_15}. One major advantage of using 2D box traps is that the dynamics of phase-separation depends only on the interplay between kinetic and interaction energy and one can encounter a novel phase-diagram which would otherwise be non-existent due to trap inhomogeneity. Finally, the 2D planar configuration is more suitable to access local densities fluctuations with respect to a 3D system, in which most of the time only column density can be measured. 

This sets the stage for our current work. Our goal is to extend the investigation to the beyond mean-field level and explore the case of a uniform 2D Bose-Bose mixture occupying two different hyperfine states and satisfying the miscibility condition at zero temperature. 
As a first step, we calculate the phase-diagram as a function of temperature and inter-species interaction strength and identify the miscible and immiscible regions using the mean-field Hartree-Fock (HF) theory~\cite{ohberg_98,schaeybroeck_13}. By treating the atoms in the two components as up and down spin states of spin-$1/2$ particles, the transition from a miscible to an immiscible mixture can be seen as a magnetic phase-transition. In the HF theory it occurs where the spin susceptibility exhibits a sudden discontinuity. A remarkable result of the theory is that, in the superfluid regime, temperature increase tends to favor phase separation. This result seems counter-intuitive, as thermal fluctuations usually acts against order. An example of a similar type of anomaly in the context of classical fluid is the temperature driven phase-transition in water-triethylamine mixtures~\cite{Atkins}.

As a second step, we use the Stochastic (Projected) Gross-Pitaevskii (SGPE)~\cite{blakie_08,proukakis_08} theory for the same mixtures. This formalism describes the system and its fluctuations by using noisy classical fields coupled to a thermal bath, and includes effects of thermal fluctuations both in the density and spin channels, going beyond the HF description. We calculate the density profiles of the two components at equilibrium at different temperatures and interaction strengths and we simulate the response of the system to a weak external perturbation producing a spin oscillation (i.e., an out-of-phase motion of the two components). From the spin dynamics we also obtain the velocity of spin sound waves. The SGPE simulations show evidences of the miscible-immiscible phase transition in the expected range of parameters, in qualitative agreement with the HF theory, however thermal fluctuations makes the transition much smoother, thus hindering the sharp features predicted by the mean-field theory. Only for temperature $T\leqslant 0.5 T_{\rm BKT}$, the two theories provide consistent results.


\section{Miscible-immiscible phase transition}\label{Sec.HF}

\subsection{Hartree-Fock theory}

We start from a mean-field description of the uniform 2D interacting Bose mixtures, provided by the Hartree-Fock theory. This theory has been widely used in three dimensions, to investigate the equilibrium~\cite{gerbier_04,mordini_20} properties of the mixtures at finite-temperature. At equilibrium, the chemical potential of a given component $i = \{1,2\}$ below the superfluid-normal phase transition is given by \cite{pitaevskii_16,ohberg_98,schaeybroeck_13}
\begin{equation}\label{Eq.mu}
\mu_i = g_{ii} (n_{i0} + 2 n_{iT}) + g_{12} n_{3-i}  \, ,
\end{equation}
where $n_i=n_{i0}+n_{iT}$ is the density of atoms of each component. The coupling constants for the intra-species interaction, $g_{ii}$, and inter-species interaction, $g_{12}$, are related to the $s$-wave scattering lengths $a_{ij}$ and mass $m_1 = m_2 = m$ as $g_{ij} = 4\pi \hbar^2 a_{ij} / m$.
In our work, we consider the case of symmetric mixtures with $g_{11} = g_{22} = g$ and $N_1 = N_2 = N/2$ as the number of atoms.

Although the above expression Eq.~\eqref{Eq.mu} is similar in three or two dimensions, the physical meaning behind the quantity $n_{i0}$ is formally different; while in 3D the quantity $n_{i0}$  corresponds to the condensate density, this identification no longer holds in 2D, where Bose-Einstein condensation (BEC) is ruled out at finite temperature \cite{hohenberg_65,mermin_66} and $n_{i0}$ corresponds to the \textit{quasi}-condensate density, which characterizes the suppression of density fluctuations.  According to \cite{Kagan1987, Kagan_00}, one can define 
\begin{equation}\label{Eq:def_qc}
n_{i0} = \sqrt{2 \langle \vert \hat{\Psi}_i \vert^2 \rangle^2 - \langle \vert \hat{\Psi}_i \vert^4 \rangle } \, ,
\end{equation}
where $\hat{\Psi}_i$ is the bosonic field operator and $\langle \cdots \rangle$ denotes the statistical average. For sufficiently low temperatures, the quasi-condensate is known to play the same role as the true condensate \cite{Prokofev2002}. As for the thermal component, it can be defined through the Bose distribution function $n_{iT}= {\mathcal S}^{-1} \sum_\mathbf{p} f_i({\mathbf p}) = {\mathcal S}^{-1} \sum_\mathbf{p}  [e^{\beta p^2/(2m)}z_i^{-1}-1]^{-1}$, where ${\mathcal S}$ is the area and $\beta = (k_B T)^{-1}$. This expression takes into account the mean-field effects at the level of the single-particle energy through the expression $z_i=\exp[\beta (\mu_i - 2g_{ii}n_i - g_{12} n_{3-i})]$ for the fugacity. The thermal atoms density thus takes the explicit form
\begin{equation}\label{Eq.n'}
n_{iT} = -\frac{1}{\lambda_T^{2}} \ln \left( 1-z_i \right) \, ,
\end{equation}
where $\lambda_T$ is the thermal de Broglie wavelength. It is worth noticing that Eq.~\eqref{Eq.n'} yields a divergent behavior for $\mu_i =2g_{ii} n_i + g_{12} n_{3-i}$, in accordance with the fact that BEC does not exist in 2D at finite temperature. However, in the BKT superfluid phase one can safely use expression~(\ref{Eq.mu}) for the evaluation of thermodynamic quantities.

\subsection{Thermodynamic quantities}

In order to calculate the phase diagram of the mixture, one can investigate the dynamical stability of the system. The mixture is found to be stable against density and spin fluctuations if the compressibility $\kappa_T$ and spin susceptibility $\kappa_M$ are positive \cite{landau_SP,viverit_00}. The two quantities are defined through the chemical potential Eq.~(\ref{Eq.mu}) following the thermodynamic relation $\kappa_T (\kappa_M) = \left[ \partial (\mu_1 \pm \mu_2) / \partial (n_1 \pm n_2) \right]^{-1}$.
In the particular case of the symmetric mixtures considered in this work, the compressibility as well the susceptibility assume the simple form
\begin{equation}\label{Eq.kappaM}
    \kappa_T = \frac{2 \kappa_T^\mathrm{sc}}{1 + g_{12} \kappa_T^\mathrm{sc}} \, , \quad \kappa_M = \frac{2 \kappa_T^\mathrm{sc}}{1 - g_{12} \kappa_T^\mathrm{sc}} \, ,
\end{equation}
where we have introduced the isothermal compressibility for the single-component Bose gas:
\begin{equation}\label{Eq.kappaT_sc}
\kappa_T^\mathrm{sc} = \frac{1}{g} \frac{1-g \beta \left(e^{\frac{\beta}{2}gn_0}-1 \right)^{-1}/\lambda_T^2}{1-2 g \beta \left(e^{\frac{\beta}{2} gn_0}-1 \right)^{-1}/\lambda_T^2} \, ,
\end{equation}%
where $n_0 = n_{10} + n_{20}$ is the total quasi-condensate atom density. At $T=0$, Eq. \eqref{Eq.kappaT_sc} yields the well known result $\kappa_T^\mathrm{sc}= 1/g$, and the spin susceptibility \eqref{Eq.kappaM} reduces to $\kappa_M= 2/(g-g_{12})$, revealing its large increase near the miscible-immiscible phase transition occurring for $g_{12}=g$ \cite{bienaime_16,fava_18}. At finite temperature instead, the miscible mixture is stable under the condition $1 - g_{12} \kappa_T^\mathrm{sc} > 0$. Remarkably, the isothermal compressibility in Eq. \eqref{Eq.kappaT_sc} increases at finite temperature, compared to the value predicted at $T=0$. This behavior is the direct consequence of exchange effects \cite{pitaevskii_16,pethick_08}, which is responsible for both the factor 2 in the denominator of Eq. \eqref{Eq.kappaT_sc}, as well as the temperature dependence of the chemical potential in Eq. \eqref{Eq.mu}.
\par
\begin{figure}[t]
\begin{center}
\includegraphics[width=0.9\columnwidth]{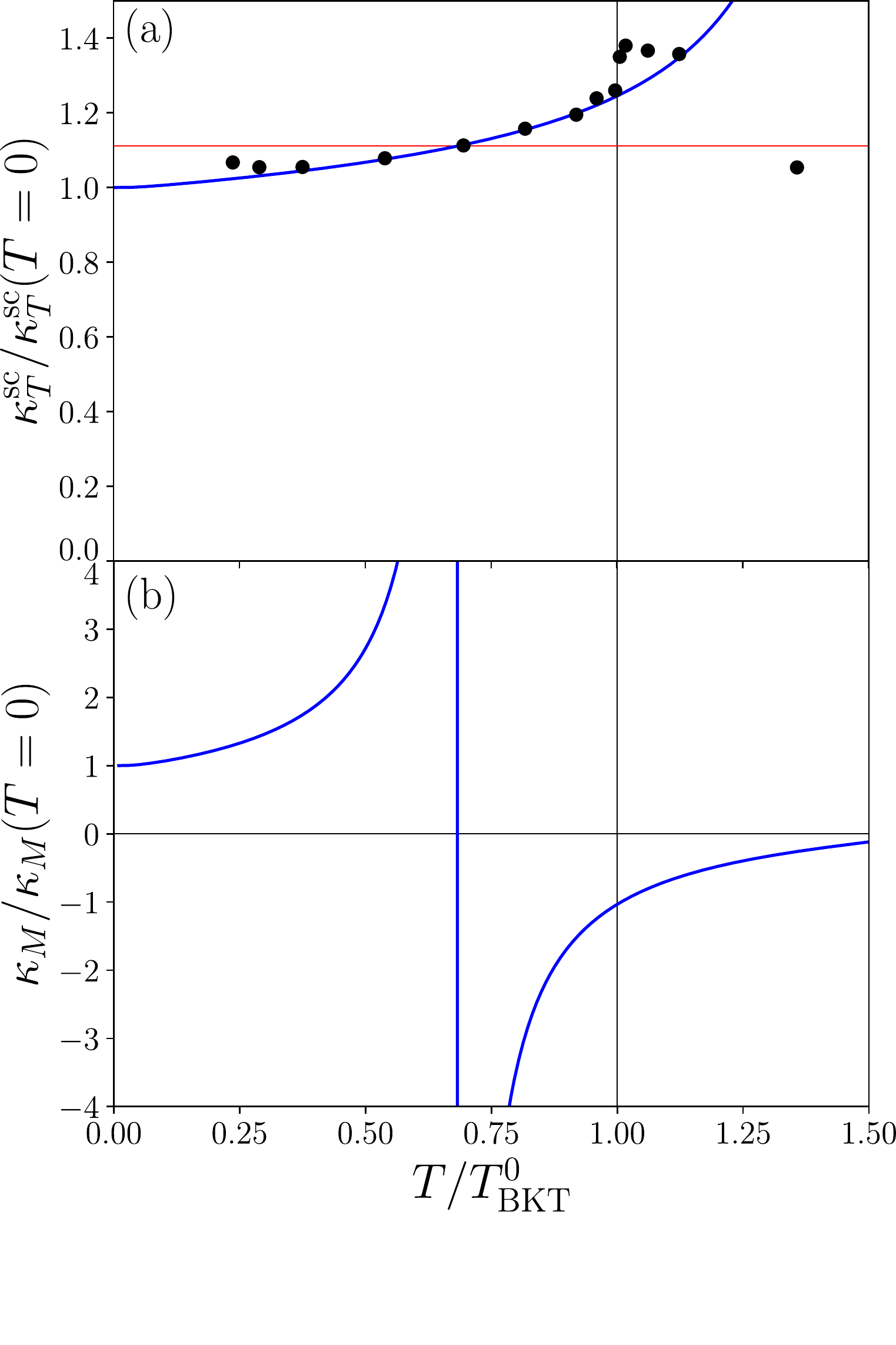}
\caption{(a) Isothermal compressibility $\kappa_T^\mathrm{sc}$ of a single-component uniform 2D Bose gas with $mg/\hbar^2=0.095$, normalized to its $T=0$ value. Temperature is normalized to the BKT critical temperature $ T_{\mathrm{BKT}}^0$ (see main text). The blue solid line is the prediction of the HF theory Eq.~(\ref{Eq.kappaT_sc}), while the black circles are results from the universal relations of Ref.~\cite{Prokofev2002}. (b) Spin susceptibility $\kappa_M$, Eq.~(\ref{Eq.kappaM}), of a two-component mixture with interaction parameters $g_{11}=g_{22}=g$, $mg/\hbar^2=0.095$ and $g_{12} / g =0.9$. The quantity  $\kappa_T^\mathrm{sc} = 1/g_{12}$ is also plotted as the red horizontal line in panel (a); the magnetic instability occurs at the point where it crosses the blue line.}  
\label{fig:thermo2D}
\end{center}
\end{figure}

In Fig.~\ref{fig:thermo2D}(a), we show the single-component compressibility $\kappa_T^\mathrm{sc}$ calculated from Eq.~\eqref{Eq.kappaT_sc}, using a typical interaction parameter $m g/\hbar^2 = 0.095$ \cite{ville_18}. For comparison, we also show the results obtained from the universal relation. The latter predicts that the equation of state of a single-component Bose gas of density $n^\mathrm{sc}$ in the vicinity of the critical density, $n_\mathrm{BKT}$, depends on a single variable $X$ related to the interaction $g$ and to the reduced chemical potential $\beta \mu$ according to $n^\mathrm{sc}-n_\mathrm{BKT}=f(X)$. The universal function $f$ has been evaluated in Ref.~\cite{Prokofev2002} using classical Monte-Carlo simulations. In addition, universal relations also provide a prediction for the BKT superfluid transition temperature of a single-component uniform Bose gas: $T_\mathrm{BKT}^0 = 2\pi \hbar^2 n^\mathrm{sc} /[ m k_B \ln (380\hbar^2 / mg)]$ \cite{prokofev_01}.  It is worth noticing that the Hartree-Fock theory is able to capture the qualitative behavior of the isothermal compressibility below $T_\mathrm{BKT}^0$, with its characteristic increase when $T$ approaches the BKT transition temperature, although it does not predict the typical peak just above the critical point. 

Our result for the spin susceptibility of the mixture, Eq.~\eqref{Eq.kappaM}, is reported in Fig.~\ref{fig:thermo2D}(b) for $g_{12}/g = 0.9$. Temperature is normalized to $T_\mathrm{BKT}^0$ which for the mixture we define according to
\begin{equation}\label{Eq:T_BKT}
    T_\mathrm{BKT}^0 = \frac{2\pi \hbar^2}{m} \frac{N}{2\mathcal{S}} \ln \left( \frac{380 \hbar^2}{mg} \right) \, .
\end{equation}
Since the inter-species interaction is expected to play a little role for the critical point of the BKT superfluid phase transition~\cite{karle_19,kobayashi_19}, the value of $T_\mathrm{BKT}^0$ for a symmetric miscible mixture is close to the actual value of BKT critical temperature; it significantly underestimates it only in the case of full phase separation. Remarkably, the instability condition $\kappa_T^\mathrm{sc} = 1/g_{12}$, which is graphically represented in panel (a) by the intersection between the compressibility curve and the red horizontal line, takes place at a temperature well below the BKT transition, and the spin susceptibility diverges at $T_M \simeq 0.68 T_\mathrm{BKT}^0$. 

\begin{figure}[t]
\begin{center}
\includegraphics[width=1.0\columnwidth]{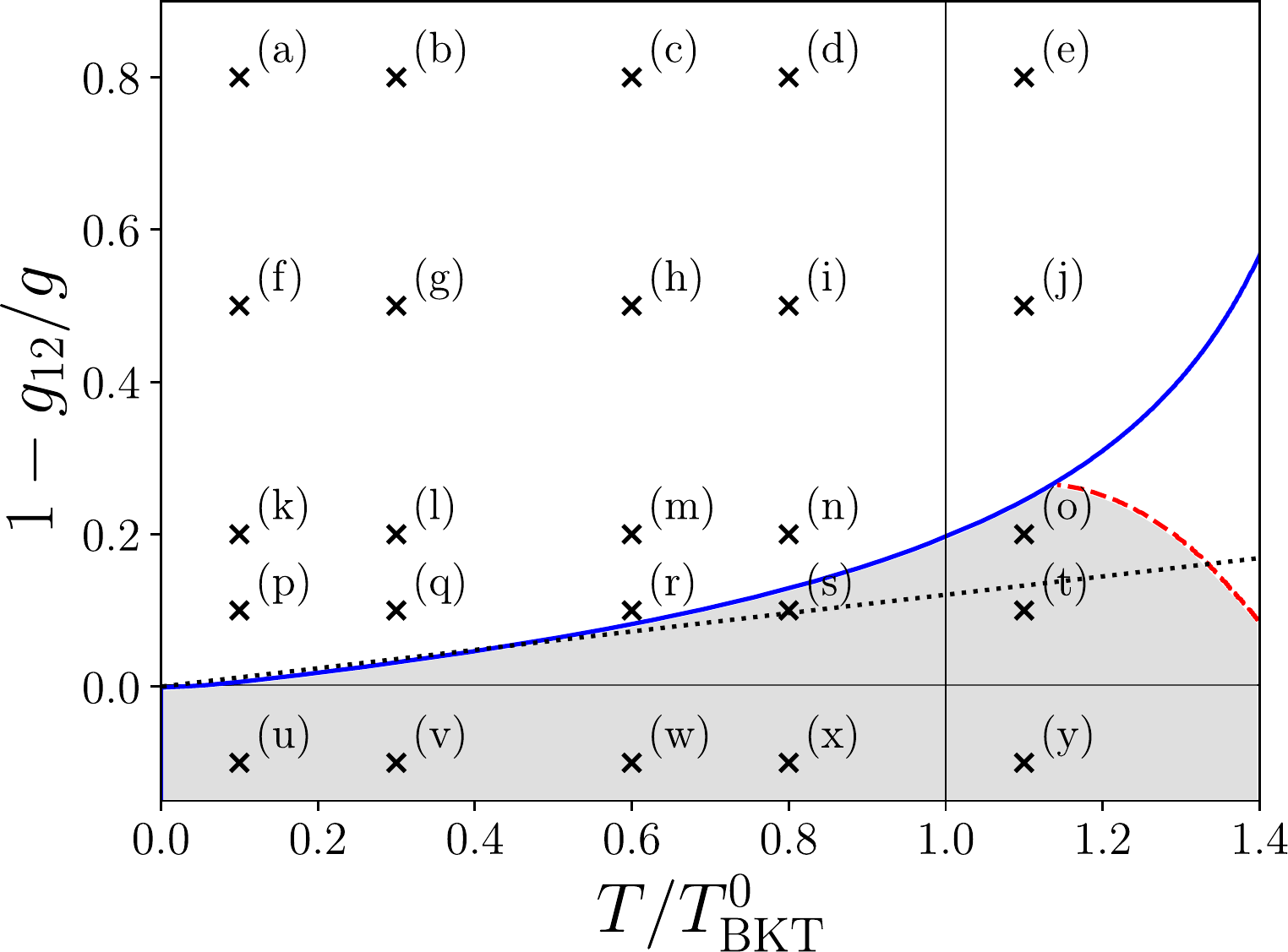}
\caption{Phase diagram of a 2D two-component Bose gas with $g_{11}=g_{22}=g$ and $mg/\hbar^2=0.095$. The blue solid and the black dotted lines are the HF predictions for the magnetic instability temperature $T_M$ evaluated from the condition $\kappa_T^\mathrm{sc} = 1/g_{12}$ and the small $g$ expansion Eq.~\eqref{Eq.TM2D}, respectively. The red dashed line is an estimate of the location of the second magnetic instability, above $T_{\rm BKT}^0$, obtained from universal relations, namely, by interpolating the data points for the equation of state given in Ref.~\cite{Prokofev2002}. According to the HF theory, the mixture is miscible above the blue and red lines, while it is dynamically unstable against phase separation  in the gray area. The array of markers correspond to the parameters used in SGPE numerical simulation and the letters serve as a guide to read Figures \ref{density} and \ref{fig:M_finiteT}.} 
\label{fig:phaseDiag2D}
\end{center}
\end{figure}

\subsection{Mean-field phase-diagram}

The pole of the spin susceptibility $1-g_{12}\kappa_T^\mathrm{sc}=0$ identifies the temperature $T_M$ at which a magnetic instability occurs. We can calculate its location for different values of the interaction strengths. The result is shown as a blue line in Fig.~\ref{fig:phaseDiag2D}. It is worth mentioning that the decrease of $\kappa_T^\mathrm{sc}$ above $T_\mathrm{BKT}^0$ predicted by the universal relations in Fig.~\ref{fig:thermo2D}(a) suggests that the instability condition characterizing the magnetic phase transition should also be satisfied above the BKT transition. We estimate its location by interpolating the data points for the equation of state given in Ref.~\cite{Prokofev2002} and we plot the corresponding $T_M$ as a red dashed line. In the interaction vs. temperature phase diagram of Fig.~\ref{fig:phaseDiag2D}, the mixture is expected to be dynamically unstable against phase separation in the gray area below the blue or red lines.

We can also derive an analytical result for the temperature $T_M$  by using the small-$g$ expression for the single component compressibility: $1 / \kappa_T^\mathrm{sc} \simeq g(1-4\mathcal{D}_0^{-1})/(1-2\mathcal{D}_0^{-1})$, with the phase space density $\mathcal{D}_0=\lambda_T^2 n_0$. This simple expression allows one to see how the isothermal compressibility reflects the universal behavior of a 2D Bose gas. Indeed, $\kappa_T^\mathrm{sc}/\kappa_T^\mathrm{sc}(T=0)$ does not explicitly depend on the value of the coupling constant $g$ and, 
in the regime $T \ll T_\mathrm{BKT}^0$, one has $\mathcal{D}_0 \simeq \lambda_T^2 n  \gg 1$ with $n=n_1+n_2$, thus obtaining the simple estimate
\begin{equation}\label{Eq.TM2D}
\frac{m k_B T_M}{\pi\hbar^2n} \simeq 1 - \frac{g_{12}}{g},
\end{equation}
valid for $gn \ll k_B T_\mathrm{BKT}^0$ and $1-g_{12}/g \ll 1$. This expression is shown as a black dotted line in Fig. \ref{fig:phaseDiag2D}, which nicely agrees with the behavior of the pole of the spin susceptibility (blue line) at low $T$. 

The above results for the magnetic instability of the binary mixture suggest the occurrence of a first order phase transition, the value of $T_M$ corresponding to the spinodal temperature above which the unpolarized uniform configuration of the mixture is dynamically unstable. The actual  transition to a demixed configuration is then expected to take place at smaller values of the temperature and could be identified by comparing the free energy of the uniform unpolarized configuration  with the one of the phase-separated configuration. In a similar way, the phase diagram for the 3D Bose mixtures has been obtained in Ref.~\cite{ota_19}, by means of the Popov theory. The equilibrium configuration in the new phase-separated phase was found to be characterized by a full space separation of the Bose-Einstein condensed components of the two atomic species, their thermal components remaining instead mixed, but with a finite magnetization. However in 2D, the calculation of the free energy, as well as the characterization of the new phase can not be assessed within the actual HF theory. In fact, any mean-field framework based on the notion of the quasi-condensate is expected to fail above the BKT transition point, where vortex proliferation destroys quasi-long range order, and the quasi-condensate becomes ill defined. We therefore need a reliable theoretical framework, which allows for the description of the 2D binary mixtures in both the superfluid and normal regimes.

\section{Stochastic Gross-Pitaevskii theory}\label{model}
 
In the following we characterise the behaviour
of the weakly interacting Bose-Bose mixture by using the coupled Stochastic (Projected) Gross-Pitaevskii equations~\cite{stoof_01,proukakis_06,proukakis_08,blakie_08,bradley_08,cockburn_09,su_11,rooney_13,davis_13,berloff_14,brewczyk_2007,gallucci_16,kobayashi_16,ota_18}
\begin{eqnarray}
	i\hbar\frac{\partial}{\partial t}\psi_i({\bf x},t)&=& {\mathcal {\hat P}}\bigg\{
		(1-i\gamma)\bigg[-\frac{\hbar^2 \nabla^2}{2m_i} + V({\bf x})
		+ g|\psi_{i}({\bf x},t)|^2 \nonumber\\
		&+&g_{12}|\psi_{3-i}|^2 - \mu_i\bigg]\psi_i({\bf x},t) + \eta_i({\bf x},t)\bigg\},
\label{cspgpe}
\end{eqnarray}
where ${\bf x} \equiv (x,y)$ are the Cartesian coordinates, the operator $\nabla^2$ is the Laplacian in 2D, 

\begin{equation*}
V(x,y)= 
\begin{cases}
0 \ \ {\rm for} \ \{x,y\}\in (0,L_x)\times(0,L_y) \\ 
\infty \ \ {\rm elsewhere},
\end{cases}
\end{equation*}
is the confining box potential, and the two complex functions $\psi_{i}({\bf x},t)$ are the ``classical" fields ($c$-fields) accounting for the macroscopically occupied low-energy modes of each component of the gas (labelled by the index $i \in \{1,2\}$) subject to random thermal fluctuations. The $c$-fields represent the coherent region of the energy spectrum, which includes a large but finite number of low-lying modes up to an energy cutoff $\epsilon_{{\rm cut}}^{i}$. The energy cut-off is chosen as~\cite{blakie_08,proukakis_08,rooney_10,comaron_19,fabrizio_18,liu_20}
\begin{equation}
	\epsilon_{{\rm cut}}^{i} = k_{\rm B} T\ln 2 + \mu_i \, .
\end{equation}
This choice guarantees that the mean occupation of the modes below $\epsilon_{{\rm cut}}^{i}$ is of order $\sim 1$ or larger,
if one uses the single-particle Bose-Einstein distribution as an estimate.
Note that the actual value of the cut-off is not crucial and one should only ensure that it belongs to a 
reasonable range. The same choice has also been earlier used in our group to corroborate experimental results for single 
component condensates in similar configurations~\cite{ota_18,comaron_19,fabrizio_18}. The projector $\mathcal {\hat P}$ maintains the $c$-fields within the coherent region at each step of the numerical solution. The modes above the cut-off represent the incoherent region of the energy spectrum; it is the source of a stochastic Gaussian random noise which satisfies the following fluctuation-dissipation theorem 
\begin{equation}\label{Eq:Gaussian-noise}
	\langle \eta_i({\bf x},t) \eta_j^*({\bf x}',t')\rangle = 2\hbar \gamma k_{\rm B} T  \delta({\bf x} - {\bf x}')\delta(t-t')\delta_{ij} \, ,
\end{equation}
where $\langle \cdots \rangle$ denotes the averaging over different noise realizations. The amount of coupling between the coherent and incoherent regions is fixed by the parameter $\gamma$, which accounts for the thermal equilibration rate. In this work, we choose $\gamma= 0.01$, which is the same of Ref.~\cite{ota_18}. Similar values were also used in  \cite{comaron_19} and \cite{liu2018}; in the latter case, the parameter $\gamma$ was optimized to reproduce typical experimental growth rates of single component condensates in 3D. The value of $\gamma$ only affects the rate at which the system reaches equilibrium, but it does not play any role in describing the equilibrium properties. It is worth mentioning that in SGPE individual results obtained with independent noise realizations are equivalent to the individual results obtained from independent experimental runs; due to the random nature of the noise, the outcomes of each noise realizations  will differ  from  one  another  as  is  the  case in experiments. Furthermore, since SGPE describe a grand-canonical ensemble, the number of atoms is not a conserved quantity and one has to use the chemical potential in order to normalize the number density to the desired value. The density of the atoms in the $c$-fields is $n_i({\bf x},t)=|\psi_i({\bf x},t)|^2$. An energy-damped or canonical SGPE approach~\cite{rooney_15,anglin_97,gilz_11} might also be used as an alternative in order to avoid issues related to number conservation. However, due to the equivalence of canonical and grand-canonical ensemble, both variants of SGPE are expected to render similar physics for the miscible phase at equilibrium. 

In order to perform simulations which are meaningful for feasible experiments, we choose our parameters compatible to those of the ongoing experiments~\cite{gaunt_13,ville_18}, with confinement in rectangular box potential in the $x$-$y$ plane and harmonic trap in the $z$-direction. The  confinement along $z$ is sufficiently strong to freeze all degrees of freedom in that direction. In this work, we consider a uniform 2D hard-wall square box of dimensions $L_x \times L_y = (50 \times 50)\mu$m. The frequency of the harmonic confinement in the experimental geometry, $\omega_z$, can be used to relate the actual 3D $s$-wave scattering length $a_{ij}$ of the atoms in different hyperfine levels to the 2D coupling constants $g$ and $g_{12}$ used in this work, related by $g_{ij} = \sqrt{8\pi}a_{ij}/a_z$ such as to ensure that our simulations correspond to realistic configurations. Here, $a_z = \sqrt{\hbar/m \omega_z}$ is the oscillator length along $z$. A typical atom number in our simulation is $1.8 \times 10^4$ in each component, which corresponds to  $\mu_0/k_{\rm B} = 4.8$~nK and $T_{\rm BKT}^0 = 33.25$~nK, if $\omega_z= 2 \pi \times (1500\ {\rm Hz})$. 

\begin{figure*}
\includegraphics[width=0.9\linewidth]{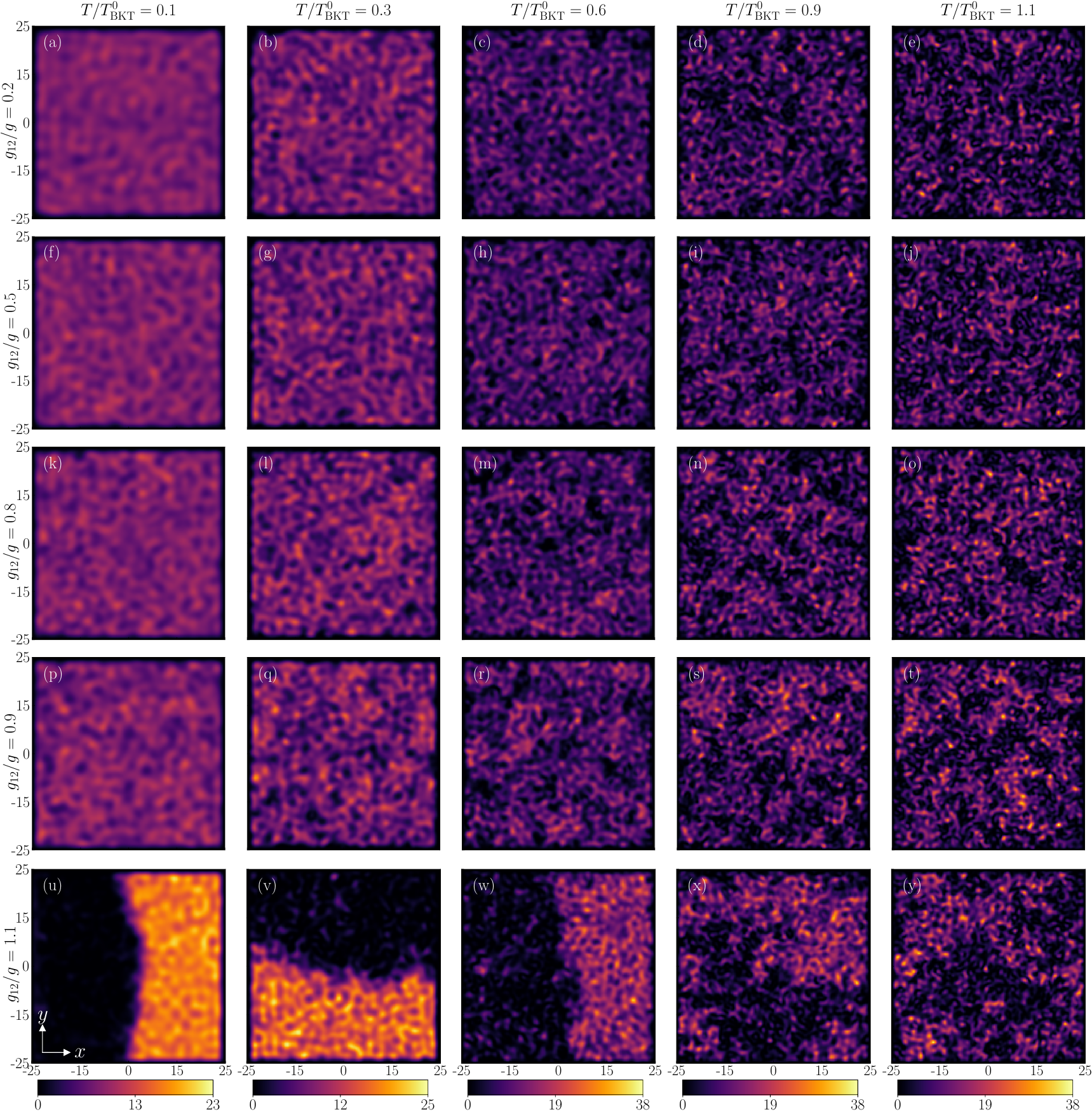}
\caption{Equilibrium $c$-field density profiles of the first component $n_1$ obtained from single runs of SGPE simulations at different inter-species interaction strengths and temperatures in an uniform 2D hard-wall square box of dimensions $L_x \times L_y = (50 \times 50)\mu$m. Colorbar represents the density of atoms in units of $\mu$m$^{-2}$, where black denotes the absence of atoms, and bright yellow as the maximum density. The values of $T/T_{\rm BKT}^0$ and $1-g_{12}/g$ in each panel are represented in Fig.~\ref{fig:phaseDiag2D} by the point associated to the same letter. }
\label{density}
\end{figure*}

\section{Equilibrium density profiles}
\label{Sec.Equilibrium}

We obtain the equilibrium configurations at a given temperature $T$ by numerically propagating Eq.~(\ref{cspgpe}) in real-time starting from purely random $c$-fields until equilibrium is reached, with the constraint $\mu_1 = \mu_2 = \mu$. For a finite-size uniform 2D Bose-Bose mixture one can choose the value of the chemical potential from the relationship $\mu \approx (1 + g_{12}/g)\mu_0$ obtained from Eq.~\eqref{Eq.mu} at zero temperature, $\mu_0$ being the chemical potential of a single component Bose gas. We have verified, using the universal relations of Ref. \cite{Prokofev2002}, that for a single-component 2D Bose gas, the temperature dependence of the chemical potential is weak enough, so that one can use its $T=0$ value in the whole region of temperature of our interest.  In our numerical simulations, an equilibrium configuration is assumed to be reached at a time when the number of atoms in each component saturates around a mean value, and the fluctuations in the number of atoms 
around the respective time average is small (typically  within 8\% and 3\% for the highest and lowest temperatures, respectively)\cite{cockburn_10}.

The density profiles obtained from typical SGPE simulations with different values of $g_{12}/g$ and $T/T_{\rm BKT}^0$ are shown in Fig.~\ref{density}. Each panel represents the density $n_1({\bf x},t)$ obtained in the evolution of a mixture starting from a single noise realization. The density $n_2({\bf x},t)$ is similar, except that it is larger when $n_1$ is smaller, in such a way that the total density is almost uniform in the box, apart from thermal fluctuations. The first two rows correspond to values of the interaction parameters for which HF theory predicts full miscibility at all temperatures. Conversely, the lowest row falls into the gray area of Fig.~\ref{fig:phaseDiag2D} where the mixture is expected to be immiscible. The SGPE density profiles are consistent with these predictions, with phase separation clearly visible in the last row, especially at low $T$. 

The third and fourth rows of Fig.~\ref{density} are those where signatures of the miscible-immiscible phase-transition are expected to appear. According to HF theory, the parameters of panels (o), (s) and (t) are such that the mixture should be phase-separated, though it is miscible at $T=0$ for the same interaction strength. Actually, SGPE shows random patterns, with archipelagos of atoms, albeit marred by fluctuations in the equilibrium density profiles, having a close resemblance to the density profiles for $g_{12}/g = 1.1$, in panels (x) and (y), where the immiscibility tends to get suppressed because of thermal fluctuations.

A caveat of the SGPE simulations is that, depending on the interaction parameters, the initialization of chemical potential $\mu$ can have dramatic effects on the composite system at equilibrium. In particular, with equal $\mu_i$ 
and $g_{12}/g>1$ and for certain noise realizations, due to the grand-canonical nature of the SGPE, the number of atoms ($N_1$ or $N_2$) of one of the components may decrease considerably in order to lower the total energy. Anomalous trajectories, with large population imbalance $(N_1-N_2)/(N_1+N_2)$, are thus manually rejected.  For the configurations shown in Fig.~\ref{density} and those used in ensemble averages, the imbalance is always small, being significant only in the immiscible region of the phase diagram and at low temperature, corresponding to the panels (u), (v), and (w) of Fig.~\ref{density}, where the relative imbalance is of the order of $0.1$ or less; in all the other cases, it is negligible, i.e, $\sim 10^{-3}$ or smaller. 
Furthermore, when performing configuration averages in the immiscible region, one has to pay attention also to the different spatial orientations of the density patterns, which may hinder the signatures of phase-separation in the average. These cautions are not needed in the miscible configurations, and they are also irrelevant in our dynamic simulations where, as we will see later, the coupling to the bath is switched off, thus ensuring the subsequent number conservation.

\begin{figure}[!hbtp]
\includegraphics[width=1.0\linewidth]{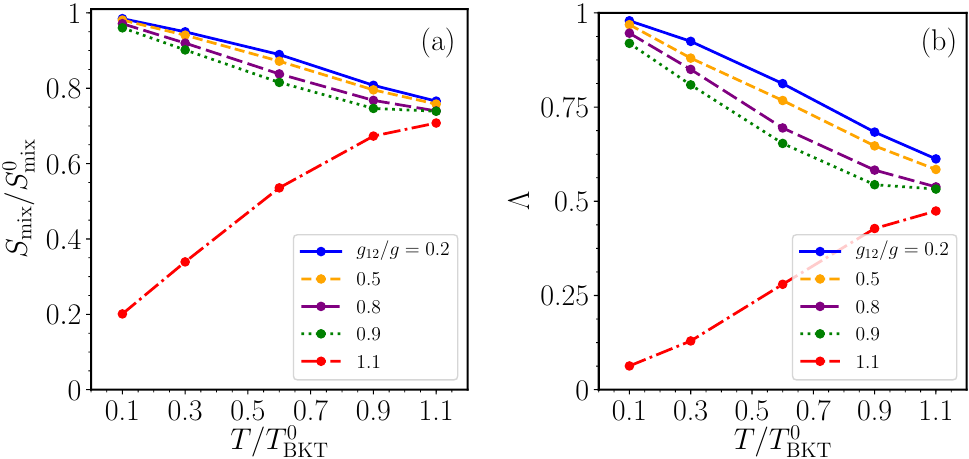}
\caption{(a) Mixing entropy, $S_{\rm mix}$, defined in Eq.~(\ref{entropy}) and normalized to its maximum value, $S_{\rm mix}^0  = N_{\rm cell}\ln 2/2$. (b) Overlap integral (\ref{overlap}). Both quantities are calculated at equilibrium and averaging over ${\mathcal N}$ noise realizations for different values of temperature and inter-species interaction strength.}
\label{meov}
\end{figure}

From the equilibrium density profiles we can also calculate the mixing entropy and the overlap integral. To obtain the mixing entropy we divide the 2D box in $N_{\rm cell}$ cells and compute the number of atoms $n_{1,j}$ and $n_{2,j}$ in the cell $j$, with $j=1,2, ..., N_{\rm cell}$ for an individual SGPE realization. For the present work $N_{\rm cell}$ equals the number of grid points used for numerical discretization. Then we use the definition  
~\cite{camesasca_06,brandani_13,Penna_2019}
\begin{eqnarray}
S_{\rm mix} &=& -\frac{1}{2}\sum_{j=1}^{N_{\rm cell}}\Bigg[\frac{n_{1,j}}{n_{1,j}+n_{2,j}}\ln\bigg({\frac{n_{1,j}}{n_{1,j}+n_{2,j}}}\bigg) \nonumber\\
&+& \frac{n_{2,j}}{n_{1,j}+n_{2,j}}\ln\bigg({\frac{n_{2,j}}{n_{1,j}+n_{2,j}}}\bigg)\Bigg].
\label{entropy}
\end{eqnarray}
The result is given in Fig.~\ref{meov}(a). Where $S_{\rm mix}$ has been averaged over ${\mathcal N = 40}$ realizations. In the low temperature limit the mixing entropy is large for $g_{12}/g < 1$ (miscible mixture) and small for $g_{12}/g > 1$ (immiscible), while at large temperature all the curves tends to the same value.
A similar behavior is also found in the temperature dependence of the overlap integral defined as 
\begin{equation}
\Lambda = \frac{2\int n_1 ({\bf x}) n_2 ({\bf x}) \ d{\bf x}}{\int (n_1^{2} ({\bf x}) + n_2^{2} ({\bf x}))  \ d{\bf x}},
\label{overlap}
\end{equation}
shown in panel Fig.~\ref{meov}(b). Here too, $\Lambda$ has been calculated for equilibrium density profiles obtained from a single SGPE simulation, and then averaged over ${\mathcal N}$. In both cases, the purple and green line corresponding to $g_{12}/g=0.8, 0.9$, respectively tend to bend more toward lower entropy and lower overlap than the curves for smaller values of $g_{12}/g$, the difference being larger at finite $T$ than at $T=0$. This mild tendency to demixing seems to qualitatively agree with the expectations of HF theory; however, the disagreement at the quantitative level remains large, as there is no sign of sharp transitions. The convergence of all curves to a constant value at large $T$ is consistent with the presence of the Gaussian white noise source Eq.~\eqref{Eq:Gaussian-noise} in the SGPE.
Ideally, in SGPE formalism, the ensemble averaging of individual $c$-field realizations is first performed to suppress the effects of random noise, followed by the computation of different physical quantities~\cite{proukakis_08}. However, in the current work, we first calculate the pertinent observables corresponding to the equilibrium $c$-field density profiles obtained from a single SGPE run, followed by its averaging. This is essentially done to retain the signatures of immiscibility within the equilibrium solution, which, as mentioned earlier
can get washed away because of ensemble averaging at the beginning. On the contrary, the inherent presence of noise being dominant at high temperature within the individual realization of equilibrium density profiles makes the values of $S_{\rm mix}$ and $\Lambda$ saturated.


\section{Dynamical response}
\label{dynamics}

\subsection{Center of mass drift}

\begin{figure}[t]
\includegraphics[width=0.9\linewidth]{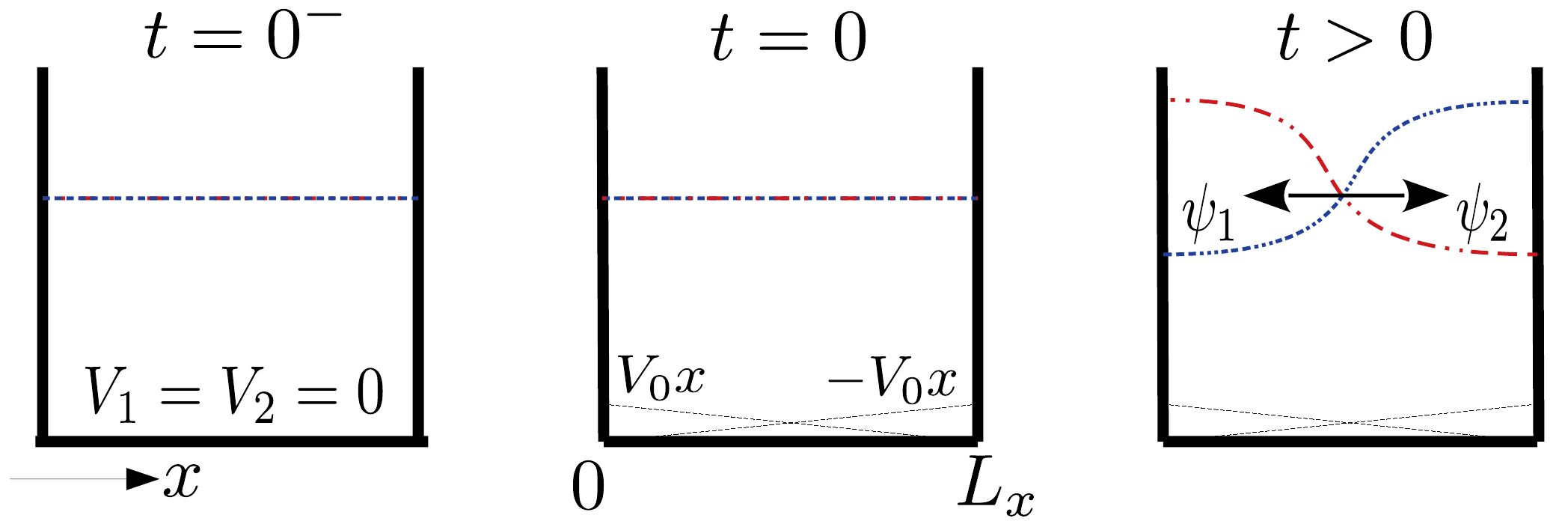}
\caption{Schematic diagram to illustrate the linear ramping of the potential to probe the center of mass response. At $t=0^-$, the equilibrium solution of the binary mixture in an uniform box is obtained through SGPE simulations. Then, at $t=0$, a positive and negative linear potential ramp of the form $V_0 x$ for respective species is switched on. The center of mass response of each of the species is then probed through the dynamics of the composite system for $t>0$.}
\label{Fig.pottilt}
\end{figure}

The miscibility of binary Bose mixtures can also be assessed within the SGPE theory by evaluating the spin dynamics. For this purpose, we follow the approach adopted in the experimental work of Ref.~\cite{bienaime_16}, and evaluate the spin center of mass oscillation by applying an external potential. We first generate the equilibrium density profiles using Eq.~(\ref{cspgpe}) at given interaction strengths and temperature as described in the previous Section. The mixture is then suddenly subjected to a weak potential tilt, acting on the two components in opposite directions. We choose a linear potential of the form $V_\mathrm{ext} = V_0 x \sigma_z \theta (t)$, where $V_0$ is the strength of the potential which is of the order $10^{-2}\mu_0$ and $\sigma_z$ the Pauli matrix (see Fig.~\ref{Fig.pottilt}).  The system is then let to evolve in absence of dissipation, i.e., with $\gamma=0$. This ensures the number of atoms to be conserved all throughout the evolution. A similar experimental protocol has been employed to measure the speed of sound in a box for a single-component Bose gas, both in 2D~\cite{ville_18} and 3D~\cite{garratt_19}.

The application of the external potential renders each species of the mixture adapt itself to a displacement along the direction of the potential minima. These bring about marked changes in the density and sound waves are emitted. From the evolution, we extract the center of mass drift along the $x$ and $y$ direction through the definition
\begin{equation}
	{\bf x}_i^{\rm cm}(t) = \frac{\int{\bf x}|\psi_i({\bf x},t)|^2 {\rm d}{\bf x}}{\int|\psi_i({\bf x},t)|^2 {\rm d}{\bf x}} \, .
	\label{com}
\end{equation}
The dynamics of the center of mass drift through Eq.~(\ref{com}) is then averaged over ${\mathcal N} = 20 \sim 40$. In what follows, we investigate in particular the spin dipole moment, given by the center of mass drift according to 
\begin{equation}
\mathbf{M}(t) = \mathbf{x}_1^\mathrm{cm}(t) - \mathbf{x}_2^\mathrm{cm}(t) \, .
\label{Mt}
\end{equation}

We notice that the method of evolving in time the states  stochastically generated at $t=0^-$, using the projected Gross-Pitaevskii equation for the classical field, has also been earlier used to model the growth of quasi-condensate on an atom chip~\cite{proukakis_06}, and is similar in essence to the truncated Wigner method for Bose condensed gases~\cite{sinatra_02}. 

\subsection{Linear Response Theory}

Before presenting the numerical results for the dynamics of the spin dipole moment we address the problem by means of linear response theory \cite{pitaevskii_16}. In particular we evaluate the dynamic spin response function of the 2D Bose mixtures using the Random Phase Approximation (RPA): \cite{zng_2009}
\begin{equation}\label{Eq.chi_M}
\chi_M (\mathbf{k}, \omega) = \frac{2 \chi (\mathbf{k},\omega)}{1 - g_{12} \chi (\mathbf{k},\omega)} \, ,
\end{equation}
with $\chi(\mathbf{k},\omega)$ as the density response function for a single-component Bose gas,
\begin{equation}\label{Eq.chi_SC}
\chi (\mathbf{k}, \omega) = \frac{\chi_{0}^0 + \chi_{T}^0 - g \chi_{0}^0 \chi_{T}^0}{1 + g  \left[ \chi_{0}^0 ( 1 -2 g \chi_{T}^0) +2 \chi_{T}^0 \right] } \, .
\end{equation}
The reference response functions for the quasi-condensate $\chi_{0}^0$ and thermal atoms $\chi_{T}^0$ are evaluated consistently within the HF description of Sec.~\ref{Sec.HF} (we consider the external perturbation along the $x$-direction $\mathbf{k}=k\mathbf{e}_x$ and the long wavelength limit $k \to 0$):
\begin{subequations}
\begin{align}
\chi_0^0(k,\omega) &= - \frac{n_0}{2m} \left(\frac{k}{\omega}\right)^2 \, , \\
\chi_T^0(k, \omega) &= - \frac{1}{(2\pi \hbar)^2} \int d^2 \mathbf{p} \frac{1}{p_x/m - \omega / k - i \delta} \frac{\partial f(\mathbf{p})}{\partial p_x} \, ,
\end{align}
\end{subequations}
where $f(\textbf{p})=[e^{(p^2/(2m)+gn_0/2)/(k_BT)}-1]^{-1}$ is the Bose distribution function of the 2D ideal Bose gas. 

We note on passing that the density response function used in Ref. \cite{ota_18} is retrieved from Eq. \eqref{Eq.chi_SC} by explicitly putting $n_0 = 0$ and replacing $2g$ by $g$, thus considering the gas as a normal gas with suppressed density fluctuations. Although such procedure is found to give a good description of the 2D Bose gas in the normal phase, in the present work one needs to properly take into account the effects of density fluctuations through the temperature dependence of the quasi-condensate, in order to describe the magnetic phase transition. The present RPA is compatible with the quasi-condensate HF theory of Sec. \ref{Sec.HF}, and one can immediately verify that the compressibility sum-rule $\chi_M(k \to 0 , \omega = 0) = \kappa_M$ yields the HF result for the spin susceptibility Eq. \eqref{Eq.kappaM}. 

In order to evaluate the linear response of the system under the perturbation described in Fig.~\ref{Fig.pottilt}, let us simplify the problem and consider a uniform mixture subjected to a sudden application of a spatially periodic potential, producing a sinusoidal modulation of wavevector $q_0 = \pi / L_x$. This essentially amounts to neglect finite-size effects coming from the hard-walls. Under such perturbation, the magnetization density starts to oscillate with a time-dependent amplitude given by \cite{Zambelli_01,Sartori_15}:
\begin{equation}\label{Eq:linear_theory}
M_x(t) = V_0 \kappa_M \left[1 - \frac{1}{\pi \kappa_M} \int d\omega \frac{\chi_M''(\omega / q_0)}{\omega} e^{i\omega t} \right] \, ,
\end{equation}
corresponding to the $x$-component of the spin dipole moment, and where $\chi_M''$ is the imaginary part of the spin response function given by Eq.~\eqref{Eq.chi_M}.

\begin{figure}[t]
\begin{center}
\includegraphics[width=0.8\columnwidth]{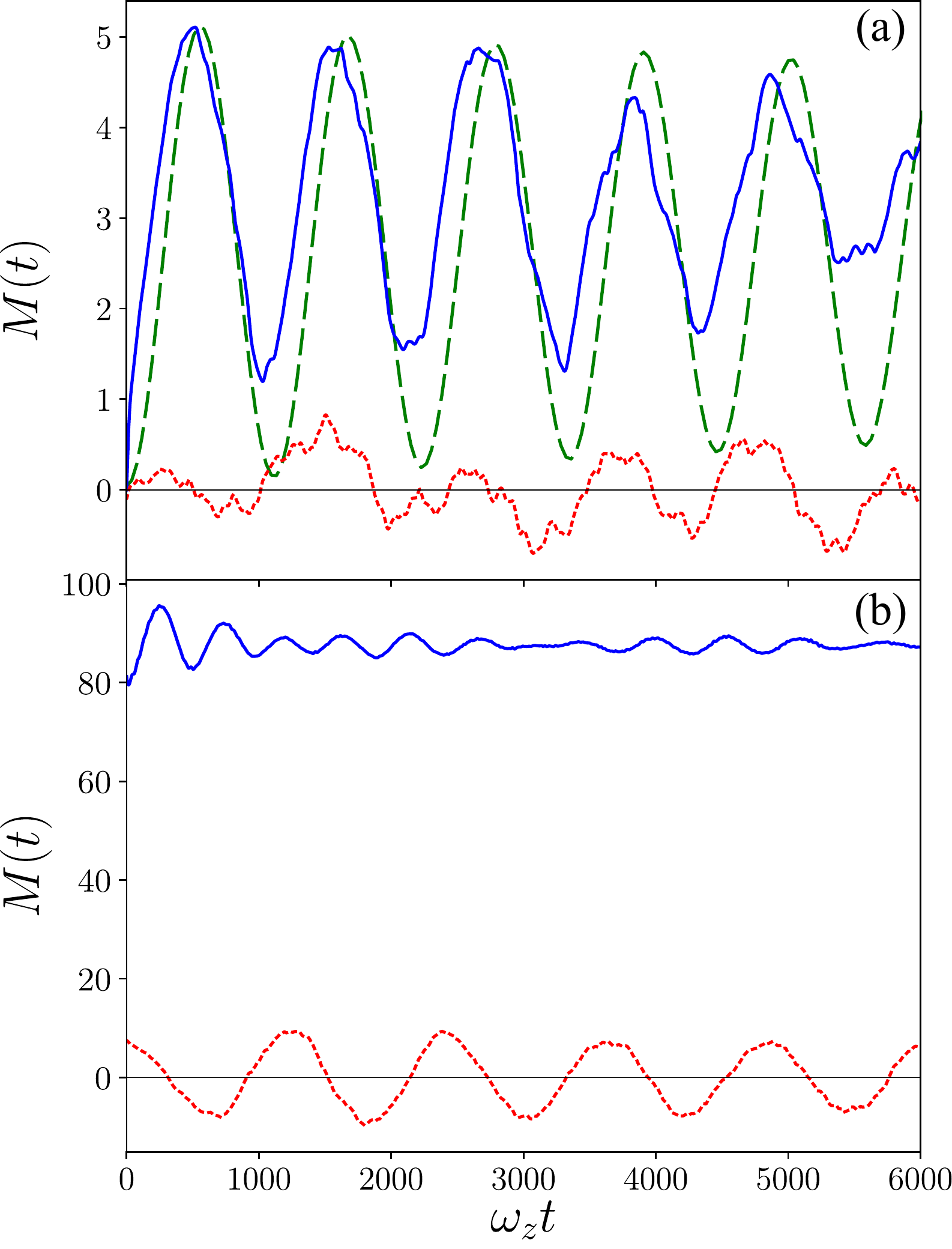}
\caption{Time evolution of the spin center of mass (\ref{Mt}) as a function of time for $mg/\hbar^2  = 0.095$, $T/T_\mathrm{BKT}^0 = 0.1$. Upper panel (a): $g_{12}/g = 0.5$. Lower panel (b): $g_{12}/g = 1.1$. The blue solid line is the oscillation in the $x$-component along which the external potential is induced, whereas the red dotted line is the oscillation in the perpendicular $y$-component. The green dashed line in the upper panel is the RPA prediction from Eq.~\eqref{Eq:linear_theory}.} 
\label{fig:M_T01}
\end{center}
\end{figure}

\subsection{Spin Oscillation Dynamics}

In Fig.~\ref{fig:M_T01}, we first show the spin oscillation dynamics $\mathbf{M}(t)$ obtained from SGPE simulation at low temperature $T/T_\mathrm{BKT}^0 = 0.1$, for two different values of coupling constant $g_{12}/g = 0.5$ (upper panel) and $=1.1$ (lower panel).
As one could expect from the zero-temperature miscibility criterion $g_{12} < g$, we find that for $g_{12}/g=0.5$, the two components are fully mixed, going back and forth in the direction where the external potential is induced (blue solid line) until reaching eventually the new equilibrium position given by the spin susceptibility (see Eq.~\eqref{Eq:linear_theory}). On the other hand for $g_{12}/g = 1.1$, the two components are already phase-separated at $t=0$; the distance between their center of mass is large along $x$ at any time, and the effect of the external potential is to induce a density oscillation in the respective component. Another signature of phase-separation is provided by the oscillation in the 
direction perpendicular to the external potential, that is, along the $y$ direction given by the red dotted 
line in Fig. \ref{fig:M_T01}. The ensuing dynamics couples both the $x$ and $y$ directions. In particular, in the $g_{12}/g = 1.1$ case phase-separation is responsible for a well-defined oscillation also in the perpendicular $y$ direction (notice the different vertical scale). For the miscible case in the upper panel of Fig. \ref{fig:M_T01}, we also show the results obtained from RPA approach Eq. \eqref{Eq:linear_theory}. In RPA, the oscillation frequency at $T \simeq 0$ is given by the spin mode of the Bogoliubov sound as $\omega = c_s q_0$, with $c_s(T=0) = \sqrt{(g-g_{12})n/(2m)}$ where $n = n_1 + n_2$ is the total density of atoms, and we find a good agreement with SGPE simulation. As for the phase-separated case we have verified that both components oscillate at the frequency of the Bogoliubov phonon mode for a single-component Bose gas occupying half area of the box.

A clear distinction between the behavior of a miscible mixture and a phase-separated gas, as in Fig.~\ref{fig:M_T01}, can only be expected at low temperature. In particular, a well defined phase-separated dynamics can be seen only in configurations like in panels (u) and (v) of Fig.~\ref{density}. The dynamical response of the system at higher $T$ and closer to the region of magnetic instability is more complex. 

\begin{figure*}[t!]
\begin{center}
\includegraphics[width=1\textwidth]{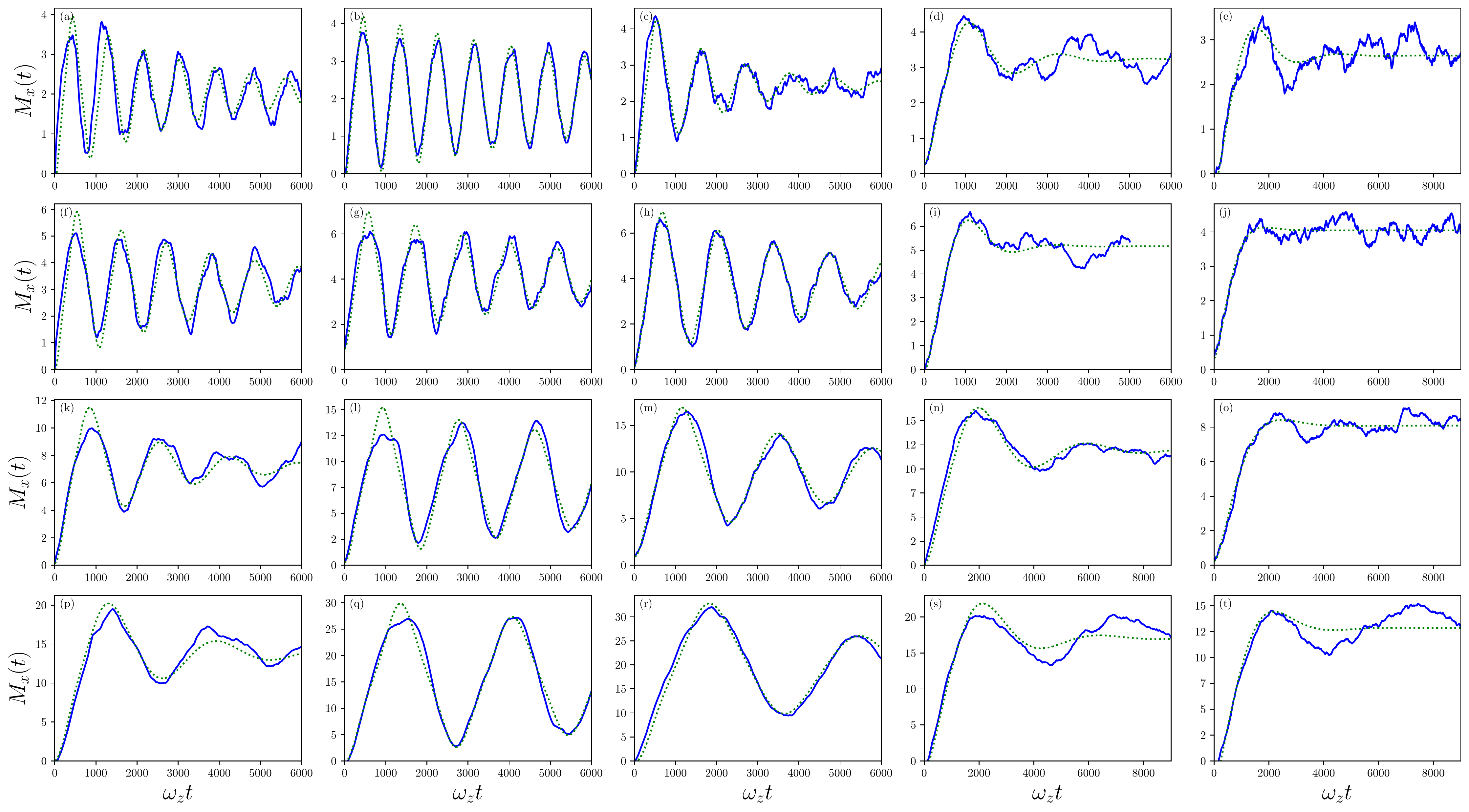}
\caption{Spin center of mass oscillation as a function of time for $mg / \hbar^2 = 0.095$. The values of temperature and interaction strength are the same as in the corresponding panels (a)-(t) of Fig.~\ref{density}. The blue solid line is the oscillation in the $x$-component (averaged over ${\mathcal N}$) along which the external potential is introduced, whereas the green dotted line is a fit using the fitting function \eqref{Eq:fit}.} 
\label{fig:M_finiteT}
\end{center}
\end{figure*}

In Fig.~\ref{fig:M_finiteT}, we show the spin center of mass oscillation obtained with SGPE simulations starting from each of the  configurations in panels (a)-(t) of Fig.~\ref{density}, followed by averaging over ${\mathcal N}$. In order to extract the frequency of the oscillation we use the expression 
\begin{equation}\label{Eq:fit}
M_x^\mathrm{fit} (t) = A_0 \left[1 - e^{-\Gamma t} \left( \cos(\tilde{\omega} t) + \frac{\Gamma}{\tilde{\omega}} \sin(\tilde{\omega}t) \right) \right] \, ,
\end{equation}
where $A_0 \propto \kappa_M$ is related to the spin susceptibility, while $\tilde{\omega} = c / q_0$ and $\Gamma$ correspond to the frequency and damping rate of the spin sound mode, respectively. Equation \eqref{Eq:fit} has been obtained by assuming a damped harmonic oscillator model for the response function $\chi_M$ in Eq.~\eqref{Eq:linear_theory}. The fitted functions are shown as green dotted lines in Fig.~\ref{fig:M_finiteT}. While we obtain good fit at low and intermediate temperature, the fitting seems to become less accurate at high temperature, where the oscillations are strongly damped and thermal fluctuations become dominant. 

\begin{figure}[t]
\begin{center}
\includegraphics[width=0.9\columnwidth]{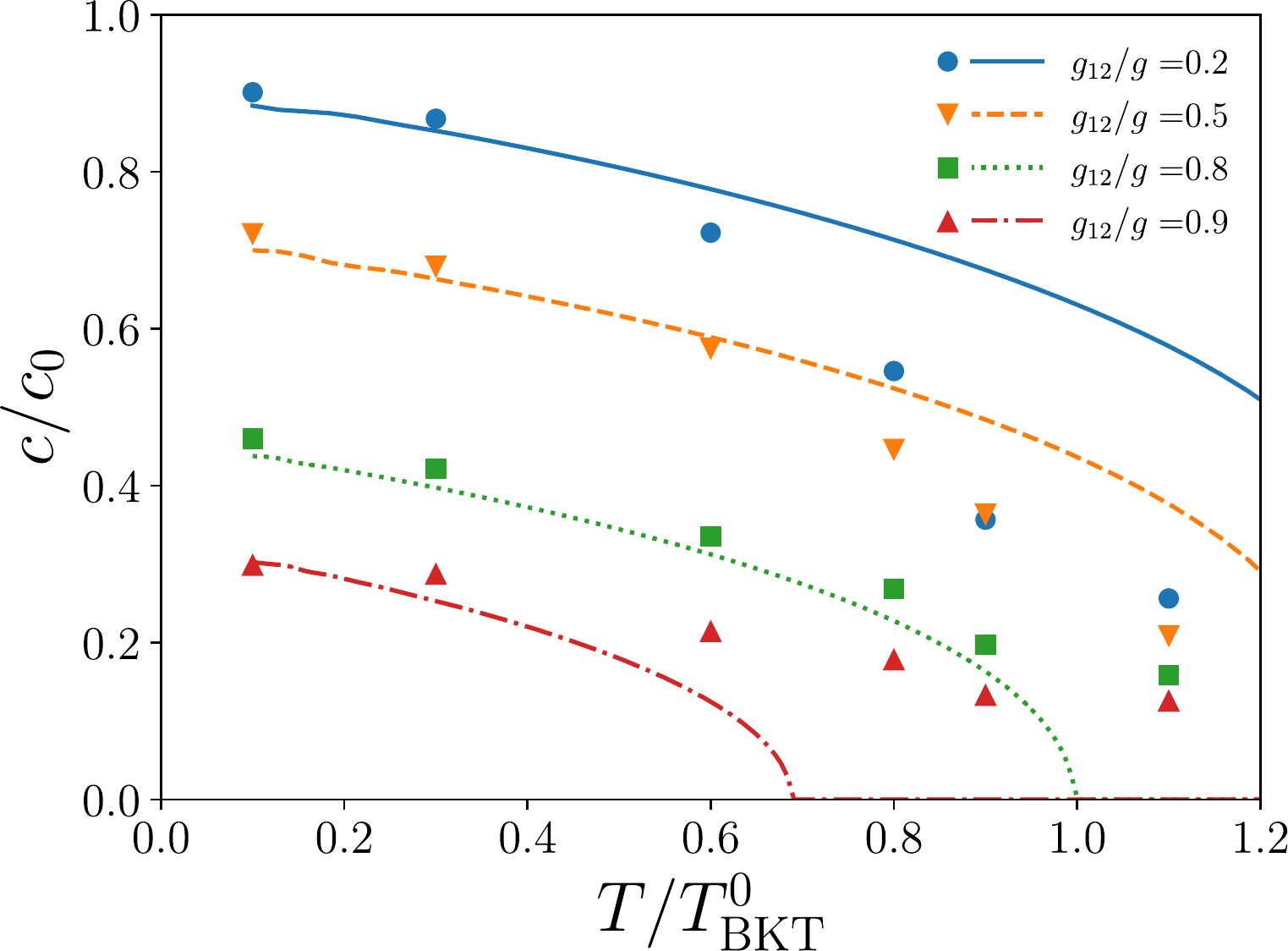}
\caption{Spin sound velocity normalized to the Bogoliubov speed of sound for a single component Bose-gas calculated for $m g / \hbar^2 = 0.095$.  From top to bottom: $g_{12}/g = 0.2, 0.5, 0.8, 0.9$. Points refer to the spin sound velocity obtained from SGPE simulation extracted using the fitting function \eqref{Eq:fit}, while lines are the RPA prediction.}
\label{fig:c}
\end{center}
\end{figure}

The values of the spin sound velocity $c$ extracted from the fit to the SGPE oscillations are shown in Figure \ref{fig:c}, where they are normalized by the single-component Bogoliubov sound $c_0 = \sqrt{gn/(2m)}$.  The solid lines correspond to the RPA predictions for the same interaction strengths. In the superfluid phase $T < T_\mathrm{BKT}^0$, the spin sound velocity is found to decrease by increasing both the temperature and the ratio $g_{12}/g$. Remarkably, RPA predicts a vanishing sound velocity at finite temperature, associated to the magnetic instability discussed in Sec.~\ref{Sec.HF}. This can be understood by recalling that the ratio between the energy weighted and inverse energy weighted moments of the dynamic response function provides an estimate for the mean excitation energy \cite{pitaevskii_16}:
\begin{equation}
c_s \simeq \frac{1}{\hbar k} \sqrt{\frac{m_1}{m_{-1}}} = \sqrt{\frac{n}{m\kappa_M}} \, .
\end{equation}
Therefore, the divergence of the spin susceptibility is responsible for the vanishing of spin sound mode. On the other hand, the speed of sound extracted from SGPE simulation do not show this kind of behavior.  This discrepancy points out that, contrary to the single-component 2D Bose gas where mean-field approach provides a qualitatively correct picture of the system below $T_\mathrm{BKT}^0
$ (see Fig. \ref{fig:thermo2D}(a) and Ref. \cite{ota_18}), the HF theory seems inadequate in describing 2D Bose mixtures, except at low $T$ where the sound velocity is indeed close to the SGPE results. This is because HF theory assumes single-particle excitation spectrum and includes only number fluctuations arising from the interaction of two thermal atoms in the same atomic species. On the contrary, SGPE includes thermal fluctuations arising from the scattering of coherent and incoherent atoms, both in the density and spin channels, as well as the ones coming from the scattering between atoms in the coherent states. 

As regards the damping rate, in general, we observe that the oscillation becomes broad and damped as one increases the temperature (that is, from left to right in each row). This is likely due to Landau damping, arising from the coupling between the collective sound mode and the thermally populated single-particle excited states, which is properly included in the SGPE theory (see Ref.~\cite{ota_18} for a discussion in the case of a single-component 2D gas). However, the quality of the fit, especially at high $T$, is not good enough to permit a quantitative analysis. In addition, in the $T \to 0$ limit, the SGPE theory, while correctly approaching the $T=0$ results for the speed of sound, is not expected to be reliable in determining the damping rate, which is more sensitive to the way in which fluctuations are included in the theory; this may also be the origin of the anomalous high damping in panels (k) and (p) in Fig.~\ref{fig:M_finiteT}.

\section{Conclusions}

We have carried out a systematic investigation of the role of thermal fluctuations in two-dimensional homogeneous bosonic mixtures of dilute atomic gases. To this end, we have employed the Hartree-Fock mean field framework complemented with beyond mean-field effects through Stochastic (Projected) Gross-Pitaevskii formalism to demonstrate the role of finite temperature on the miscibility-immiscibility phase transition. A remarkable result predicted by the mean-field theory is the existence of phase-separation induced by thermal fluctuations, occurring even for mixtures which are miscible at zero temperature ($g_{12} < g$). Analytical predictions based on mean-field HF theory are then compared with the equilibrium density solutions obtained by numerically solving SGPE at different coupling strengths and temperature. The spin dynamics of the mixture brought about by a weak external perturbation is also simulated, which shows some qualitative agreement with the mean-field predictions for temperatures well below the critical temperature for superfluidity. However, with the inclusion of beyond mean-field effects through SGPE, both the density profiles as well as spin dynamics are devoid of any sharp distinctive features near the phase transition point predicted by the HF theory. This, we believe, would also be the scenario in actual experiments. The recent availability of box like traps would be of value to corroborate these features. 

Important open issues concern the extension of the HF theory to take into account effects of thermal fluctuations in the spin channel, to clarify the observed discrepancy between the mean-field and stochastic approaches. Indeed, while HF theory is widely used in describing single-component Bose gas as well as 3D mixtures in the density channel, our work points out its failure for the investigation of spin physics. Development of beyond mean-field theory for the 2D Bose mixtures along the line of those in Refs.~\cite{andersen_02, ota_20} could provide further insight on magnetic properties of binary mixtures at finite temperatures.

\begin{acknowledgments}
We are grateful to Stefano Giorgini and Sandro Stringari for many fruitful discussions concerning the results reported 
in Sec.~\ref{Sec.HF}. We also thank Nick Keepfer for insightful discussions related to SGPE. 
A. Roy thanks S. Gautam for useful discussions. This work is supported by Provincia Autonoma di 
Trento and from INFN-TIFPA
under the project FIS$\hbar$. A.R. acknowledges support from the Italian MIUR under the PRIN2017 project CEnTraL. We 
acknowledge the CINECA award under the ISCRA initiative, for the availability of high performance computing resources and 
support.
This material is based upon work supported with GCP research credits by Google Cloud. 
\end{acknowledgments}

\bibliography{refs}{}

\end{document}